# Hypothesis Testing and Machine Learning: Interpreting Variable Effects in Deep Artificial Neural Networks using Cohen's $f^2$

Wolfgang Messner

Darla Moore School of Business, University of South Carolina, Columbia – SC, USA

Deep artificial neural networks show high predictive performance in many fields, but they do not afford statistical inferences and their black-box operations are too complicated for humans to comprehend. Because positing that a relationship exists is often more important than prediction in scientific experiments and research models, machine learning is far less frequently used than inferential statistics. Additionally, statistics calls for improving the test of theory by showing the magnitude of the phenomena being studied. This article extends current XAI methods and develops a model agnostic hypothesis testing framework for machine learning. First, Fisher's variable permutation algorithm is tweaked to compute an effect size measure equivalent to Cohen's $f^2$ for OLS regression models. Second, the Mann-Kendall test of monotonicity and the Theil-Sen estimator is applied to Apley's accumulated local effect plots to specify a variable's direction of influence and statistical significance. The usefulness of this approach is demonstrated on an artificial data set and a social survey with a Python sandbox implementation.

Wolfgang Messner is Clinical Professor of International Business at the Darla Moore School of Business, University of South Carolina (US). He received his PhD in economics and social sciences from the University of Kassel (Germany), MBA in financial management from the University of Wales (UK), and MSc and BSc in computing science and business administration after studies at the Technical University Munich (Germany), University of Newcastle upon Tyne (UK), and Università per Stranieri di Perugia (Italy). He can be reached at wolfgang.messner@moore.sc.edu.

## 1. Introduction

Since the English statistician Sir Francis Galton defined correlation and regression as statistical concepts in 1885 and Karl Pearson developed an index to measure correlation in 1895 [1], the majority of research methodologies across disciplines aim to establish additive sufficiency, that is, find variables contributing on average linearly and significantly to an outcome [2]. Using *p*-values, these concepts measure the statistical significance of effects, that is, whether they exist at all [3]. In this classic statistical approach of null-hypothesis significance testing, the evaluation is more focused on a data's fitness to a model than on using the variables to predict an outcome. But journal editors are increasingly calling on researchers to evaluate the magnitude of the phenomena being studied, rather than only their statistical significance [4] in order to build "research-based knowledge" that is "more relevant and useful to practitioners" [5]. A very informative measure is Cohen's $f^2$ [3,6], which allows both an evaluation of the model's global effect size as well as variables' local effect sizes, that is, effect sizes of each variable's substantive significance within the context of the entire model. Yet, the reporting of effect size measures remains inconsistent in extant research literature [7].

While most interesting research questions are about statistical relationships between variables, many complex processes are based on data sets that show highly nonlinear behavior. But ordinary least squares (OLS) regression analysis enforces a limited view of "straight-line relationships among equal-interval scales on which the observations are assumed to be normally distributed" [3]. This shortcoming tempts researchers to try out machine learning methods, such as deep artificial neural networks, which are able to approximate arbitrarily complex mathematical structures while being more tolerant to noise and fault at the same time [8]. Over the last few years, a variety of machine learning libraries have been developed, which allow even non-experts to apply machine learning to various problems and extract features [9,10]. However, the uptake of machine learning models in research and real-life decision making remains limited, because those models hide their underlying mechanisms in a black box [11,12]. It is difficult to assess the connection between independent (input) and dependent (output) variables with respect to strength, direction, and form. For that reason, a growing research on interpretable machine learning, referred to as explainable artificial intelligence (XAI), develops techniques for making systems interpretable, transparent, and comprehensible [13–15]. This includes attempts to reveal the contribution of input to output variables. Though, "a practical, widely applicable technology for explainable AI has not emerged yet" [16].

Researchers across disciplines are keen to compare the performance of traditional statistical inference models, such as OLS regression, with machine learning approaches in general and neural networks in particular [e.g., 17–19]. For their OLS regression models, the aforementioned publications report the variables' direction and strength of influence, statistical significance, and effect size. When using machine learning models, Cohen's $f^2$ is unfortunately not readily accessible from commonly used explainers. Researchers therefore apply novel XAI methods, such as the variable permutation algorithm, to distinguish important variables from not so important ones [20,21]. But these variable importance measures are largely visual and not equivalent to tests of statistical and substantive significance as used in inferential statistics. This lack of statistically appropriate measures unfortunately undermines the acceptance of machine learning and XAI for hypothesis testing in scientific experiments and research models.

Another set of recently proposed XAI methods include partial dependence [22], local dependence, and accumulated local effect plots [both 23]. These graphs visually show the influence of an independent variable on the model's predictions, thereby improving model interpretability. Accumulated local effect plots work better than partial dependence plots in situations where the independent variables are heavily correlated. Additionally, they are computationally more efficient [24]. But all these novel XAI methods are primarily designed to support exploratory analysis through visualization. And while "visualization is one of the most powerful interpretational tools" [22], a comprehensive statistical framework should rely on the trinity of quantification, visualization, and hypothesis testing [25].

In this paper, I aim to connect these contemporary XAI methods with the classic statistical inference measures of direction, significance, and strength. Building on Fisher's variable permutation algorithm, I provide researchers a straight-forward way of estimating Cohen's effect size $f^2$ for the input variables of a deep learning network. Additionally, I outline how a Mann-Kendall test of monotonicity [26,27] can be applied to Apley's accumulated local effect plots in order to specify an input variable's direction of influence and statistical significance. In doing so, I allow hypothesis testing with machine learning – in the full sense of classic inferential statistics. Using the deep learning open-source software library Keras 2.4.3 and the dalex model-agnostic explainer [24,28], I provide a Python sandbox implementation for other researchers to familiarize themselves with the approach. I hope to provide a pathway for replacing OLS regression methods with deep learning artificial neural networks for analyzing complex and potentially non-linear regression problems in scientific experiments and research models.

The remainder of this paper is structured as follows. In the following section, I review the differences between *Inferential Statistics and Machine Learning* and the role XAI plays in making algorithmic models interpretable. Then, I propose a new *Framework for Hypothesis Testing in Machine Learning* and give practical interpretation guidelines. In the section *Worked Examples*, I report on two applications of the framework to an artificial and a real-world data set. This is followed by a *Discussion* about the results and suggestions for future work

## 2. Inferential Statistics and Machine Learning

While all statistics starts with data, there are two distinct approaches to reach conclusions from data [29]. The first one, inferential statistics, assumes that a stochastic data model has generated the data. The model's parameters are estimated from the data. The direction and significance levels of these parameters are then used for validating it with hypothesis testing. When multivariate OLS linear, logistics, or Cox proportional hazards regressions are "fit to data to draw quantitative conclusions, the conclusions are about the model's mechanism, and not about nature's mechanism. It follows that if the model is a poor emulation of nature, the conclusions may be wrong" [29].

The second approach is algorithmic modelling, aka machine learning. Artificial neural networks are one machine learning method, which are capable of recognizing patterns without a priori knowledge of the nature of underlying relationships in the data [30]. A training algorithm learns from data fed into the artificial neural network, which is a flexible, dynamic, and nonlinear black-box model [31]. There are no assumptions that the data is drawn from a multivariate distribution. While conventional machine learning is still somewhat limited in its ability to work with raw input data, the new field of deep learning allows for design of neural network models that are composed of multiple processing layers, which learn representations of data with multiple levels of abstraction [32]. The first experimental deep neural networks were successfully trained by computing scientists in 2006, followed by commercially viable applications after 2010 [33–36]. To summarize, research in machine learning shifts the focus from data models to the properties of algorithms, which are characterized by their convergence (during iterative training cycles) and predictive accuracy (on new and hitherto unseen data). But "despite convincing prediction results, the lack of an explicit model can make machine learning solutions difficult to directly relate to existing knowledge" [37].

Explainable artificial intelligence (XAI) aims to address this downside of machine learning by developing model explanation and interpretation techniques [13–15]. Comparable to significance and effect-size tests of variables in inferential statistics, model-agnostic explainers attempt to describe how different explanatory variables contribute to an outcome. Early XAI measures have been designed to reveal the influence of the input variables on the output variables [38], such as that the values of the network weights are looked at by the general influence measure [GIM; 39], or the weights of each input variable are sequentially zeroed to gauge the network response [SZW; 40]. But in deep learning with its complex system of weight matrices connecting numerous hidden layers, the simple GIM and SZW interpretation approaches are no longer sufficient. More recently, the XAI literature has developed alternative methods, some of which are discussed in Table 1. Originating from the need to help researchers understand the black box they have created and support them in exploratory analysis, these methods still rely heavily on visualization. In the strictest sense of classical inferential statistics, they are therefore not yet suitable for hypothesis testing.

## 3. Framework for Hypothesis Testing in Machine Learning

I will now attempt to bridge the gap between algorithmic modelling and inferential statistics. Building on the XAI measures listed in Table 1, I start with outlining an algorithm to compute Cohen's effect size measure $f^2$. I continue with the Mann-Kendall monotonicity test and the Theil-Sen estimator to judge a variable's direction of influence and related statistical significance. At the end of this section, I provide guidelines for using this new framework for hypothesis testing in machine learning.

**Table 1: Extant data set-level XAI methods**

| XAI method | Description | Pros and cons |
| --- | --- | --- |
| Variable-importance measure [20] | Measures how a model's performance changes if one or a group of input variables are removed. If a variable is important, then, after permuting the variables, the performance of the model will worsen. Outputs mean RMSE loss after a certain number of permutations. | The output of the method is easy to understand and interpret. But the RMSE loss depends on the random nature of permutations and choice of the loss function. There is no quantification of statistical significance available. |
| Partial-dependence (PD) plot [22] | Shows how the expected value of model prediction behaves as a function of an input variable. Uses the average of all ceteris-paribus (CP) profiles for all observations from the data set to plot the PD profile. | Offer a simple way to summarize the effect of a particular independent variable as a visualization. Unstable for correlated input variables [24]. No quantification or statistical significance test available. |
| Local dependence profile and accumulated local effects profile [23] | Average changes in prediction (but not predictions themselves) and accumulate them over a grid. | Fast to compute and unbiased. Across the feature space, estimates have a different accuracy. Interpretation is difficult when input variables are strongly correlated, because the effect of an individual variable cannot be isolated. An interpretation across intervals is also not permissible [41] |

This table provides a literature overview of extant data set-level XAI methods applicable for deep artificial neural networks.

## 3.1 Effect Size: Cohen's $f^2$ for Machine Learning

For regression problems with both continuous independent as well as dependent variables, Cohen's $f^2$ is appropriate for calculating the model's global effect size [3,6]:

$$f^2 = \frac{R^2}{1 - R^2}$$

In this equation, $R^2$ is the proportion of variance accounted for by all independent variables relative to a model without any regressors. Much more interesting, however, is a variation of Cohen's $f^2$ for the effect of an individual variable of interest *V* [3,42,43]:

$$f_V^2 = \frac{R^2 - R_V^2}{1 - R^2}$$

In this equation, $R_V^2$ is the proportion of variance accounted for by all other variables than *V*, again relative to a model with no regressors. Thus, the numerator $R^2 - R_V^2$ reflects the variance uniquely accounted for by *V*, over and above that of all other variables. Values of 0.02, 0.15, and 0.35 are defined as thresholds for small, medium, and large values for $f_V^2$ in the behavioral sciences; values less than 0.02 point to a trivial effect [6]. Because $f_V^2$ is often not readily accessible from commonly used statistical software, it needs to be manually calculated by estimating the variance accounted for in two different regression models, one with all variables ($R^2$) and one without *V*, but all other variables ($R_V^2$).

Because of the randomness of data initialization and resource-intensive training cycles, this loco (leave-one-covariate-out) approach is not ideal for machine learning models. Instead, I propose to apply Fisher's variable permutation algorithm [20,21], which removes the effect of a variable *after* the model is built. Let *X* be the entire data set. Through randomly reshuffling the data $X_V$ associated with *V*, a modified new data set $X_V^*$ is created, on which the prediction $\hat{y}$ is calculated. If $V$ is important in the model, then, after permutation, the prediction $\widehat{y_V^*}$ should be less precise. More formally:

1. Compute $MSE(\hat{y}, X, y)$ as the mean-squared error loss function for the original data $X$;
2. Create a new $X_V^*$ by randomly permuting the $V$-th column of $X$;
3. Compute model predictions $\widehat{y_V^*}$ based on the modified data $X_V^*$;
4. Compute $MSE(\widehat{y_V^*}, X_V^*, y)$; and
5. Repeat steps 2 to 4 several times ($B$) to contain randomness and return the average $\frac{1}{B} MSE(\widehat{y_V^*}, X_V^*, y)$.

The above steps can be configured and executed in available explainers, such as dalex. Because the $R^2$ in the effect size formula refers to the population and not to sample values [3], I use all available data for $X$, that is, both the training and testing data sets combined.

Next, I need to process the output and compute the $R^2$ and $R_V^2$ for calculation of $f_V^2$. Note that the following formula are only valid because the data set $X$ is standardized, a common requirement of machine learning algorithms:

6. Compute $R^2 = 1 - MSE(\hat{y}, X, y)$ and $R_V^2 = 1 - MSE(\widehat{y_V^*}, X_V^*, y)$; and
7. Compute $f_V^2 = \frac{R^2 - R_V^2}{1 - R^2}$.

In comparison to the loco approach, the variable permutation algorithm likely overestimates a variable's effect size. While the former removes an exogeneous explainer from the model and pretends that one is not aware of it, the latter randomly changes the explainer and thereby messes with the model's explanatory capabilities. This exaggerates the mean-squared error calculation. Additionally, different models have different average responses to reshuffling of data so that the above computed effect sizes can neither be compared between models nor against the customary benchmark thresholds of small, medium, and large effects. I therefore suggest an adjustment factor, which makes the effect sizes computed with the variable permutation process comparable to the loco approach.

First, I compute a baseline $MSE(\widehat{y^*}, X^*, y)$, in which not only one variable, but all of the model's variables are randomly reshuffled. This baseline mean-squared error is a good indication of the worst possible loss function value when there is no predictive signal in the data [44]:

8. Create a new $X^*$ by randomly permuting the all columns of $X$;
9. Compute model predictions $\widehat{y^*}$ based on the modified data $X^*$;
10. Compute $MSE(\widehat{y^*}, X^*, y)$; and
11. Repeat steps 8 to 10 several times ($B$) to contain randomness and return the average $\frac{1}{B} MSE(\widehat{y^*}, X^*, y)$.

Similar to permuting a single variable, I proceed to compute the baseline $R_{base}^2$ for calculation of the baseline $f_{base}^2$:

12. Compute $f_{base}^2 = \frac{R^2 - R_{base}^2}{1 - R^2}$ with $R_{base}^2 = 1 - MSE(\widehat{y^*}, X^*, y)$.

Next, I check how inflated $f_{base}^2$ is due to the variable permutation method by relating it to the global effect size $f^2 = \frac{R^2}{1 - R^2}$ for the entire model. Accordingly, I adjust $f_V^2$:

13. Compute adjustment factor $a = \frac{f^2}{f_{base}^2}$; and
14. Adjust $f_V^2 := a \cdot f_V^2$.

To be clear, this last step adjusts the $f_V^2$ depending on the model's average response to reshuffling of data in comparison to the loco approach, making the $f_V^2$ more comparable to Cohen's original $f_V^2$. For interpretation, I therefore suggest to continue to use Cohen's original thresholds [6], that is: trivial effect size for $f_V^2 < 0.02$, small for $0.02 \leq f_V^2 < 0.15$, medium for $0.15 \leq f_V^2 < 0.35$, and large for $f_V^2 \geq 0.35$. Note that the $f_V^2$ can be larger than one, in contrast to $R^2$.

This entire approach is model agnostic, that is, it can be used for different kind of machine learning models. While the root-mean-squared-error (RMSE) loss function is more commonly used in machine learning applications, by choosing the MSE loss function instead, the resulting $f_V^2$ measure is more directly comparable with the $f_V^2$ from OLS regression models. The main disadvantage, however, is the approach's dependence on the random nature of permutations. That is, for different runs, different $f_V^2$ values will be reported. Though, randomness can be contained and accuracy enhanced through several permutation rounds (that is, repeating steps 2-4 and 8-10 several times, e.g., $B = 50$) and using the average (see steps 5 and 11 above). Self-evidently, this increases the algorithm's computation time.

### 3.2 Direction of Influence: Mann-Kendall Monotonicity Test and Theil-Sen Estimator

The effect size measure $f_V^2$ described above does not provide any information about the direction of the effect. Yet, it is quite natural to ask whether the dependent variable's values, on average, are going up, down, or staying the same. Further, it is quite natural to inquire if this trend is monotonous, that is, present across the entire range of the values of an independent variable. To answer these questions in a hypothesis-testing process, I suggest to exploit the accumulated local effect plots [23], apply the nonparametric Mann-Kendall test of monotonicity [26,27], and then quantify the effect with the Theil-Sen estimator [45].

Apley's accumulated local effect plots are visualizations that increase the interpretability and transparency of supervised machine learning models by summarizing the influence of an independent variable on the model's predictions. As opposed to the more popular (and older) partial dependence plots [22], they do not require an unreliable extrapolation with other potentially correlated independent variables [24].

The Mann-Kendall test assesses and defends the connection between independent and dependent variables by analyzing the sign of the difference between dependent variables associated with a higher independent variable and a lower independent variable, resulting in a total of $\frac{n(n-1)}{2}$ pairs of data, where $n$ is the resolution of the independent variable (that is, the number of observations available for the range of the independent variable). The Mann-Kendall test statistic is applicable in many situations, because it is "relatively effective and robust" [46], invariant to data transformation, such as a log-transformation, and does not require the data to conform to any particular distribution [47]. However, the test is not robust against serial correlation; this can lead to an over-rejection of the null hypothesis of no trend (type I error). I therefore use the Hamed and Rao modified Mann-Kendall test with a lag of three, which addresses serial correlation by applying a variance correction approach [48,49]. Because all corrections increase the type II error, the Mann-Kendall test is "best viewed as exploratory […] [and] most appropriately used to identify stations where changes are significant or of large magnitude and to quantify these findings" [50].

Next, I assess the variable's direction of influence with the Theil-Sen estimator, again based on the predictions in the accumulated local effect plot. This estimator is a robust method for determining the slope of a linear regression equation linking two variables based on the median of the slopes of all pairs of data points. Specifically, the estimator is more robust to outliers, non-normality, and heteroscedasticity than OLS linear regression. The Theil-Sen estimator computes the confidence interval via bootstrapping [45], from which I calculate the corresponding *p*-value [following the method outlined by 51].

### 3.3 Practical and Interpretation Guidelines

Using dalex as a model-agnostic explainer, I have implemented this framework for hypothesis testing as a sandbox with Keras and Python. I make it available in the Online Supplement so that other researchers can familiarize themselves with it and use it in their own work. After a neural network model has been trained on the data, the dalex explainer object creates a wrapper around the model. This wrapped model is then examined with the two steps detailed above.

For each input variable, the following measures are calculated and listed: Cohen's effect size $f_V^2$; *p*-value for the monotonicity test; slope with *p*-value for the direction of influence. Table 2 provides a suggestion on how to analyze and report these measures in a research paper. If the effect is not monotonous, continuing the exploration of the accumulated local effect plot with a change point analysis can provide additional insights [52,53].

**Table 2: Reporting guidelines for hypothesis testing**

| Monotonicity significant (*p*-value of Mann-Kendall test)? | Direction of influence (Theil-Sen slope and *p*-value)? | Effect size | | | |
|---|---|---|---|---|---|
| | | $f_V^2 < 0.02$ (trivial) | $0.02 \leq f_V^2 < 0.15$ (small) | $0.15 \leq f_V^2 < 0.35$ (medium) | $f_V^2 \geq 0.35$ (large) |
| $p \leq 0.05$ | $s < 0 \wedge p \leq 0.05$ | The independent variable exerts only a trivial effect on the dependent variable, which is monotonic, negative, and statistically significant. | The independent variable exerts a small monotonic effect on the dependent variable, which is negative and statistically significant. | The independent variable exerts a medium monotonic effect on the dependent variable, which is negative and statistically significant. | The independent variable exerts a large effect on the dependent variable, which is negative and statistically significant. |
| | $s > 0 \wedge p \leq 0.05$ | The independent variable exerts only a trivial effect on the dependent variable, which is monotonic, positive, and statistically significant. | The independent variable exerts a small monotonic effect on the dependent variable, which is positive and statistically significant. | The independent variable exerts a medium effect on the dependent variable, which is positive and statistically significant. | The independent variable exerts a large effect on the dependent variable, which is positive and statistically significant. |
| $p > 0.05$ | | The independent variable exerts only a trivial effect on the dependent variable, which is not monotonic. | The independent variable exerts a small effect on the dependent variable, which is not monotonic. | The independent variable exerts a medium effect on the dependent variable, which is not monotonic. | The independent variable exerts a large effect on the dependent variable, which is not monotonic. |

This table provides guidelines for using the machine learning framework for hypothesis testing. In addition to the write-up, the following statistics should be reported: Mann-Kendall p; Effect size $f_V^2$; Theil-Sen slope s, p.

Following established practice for reporting variable-specific effect sizes in a research report when the context is clear, I suggest to drop the *V* in the superscript, that is, simply use $f^2$ instead of $f_V^2$. In the worked examples in the next section, I will adhere to this convention.

## 4. Worked Examples

I now put the hypothesis-testing framework into action on one artificial and one real data set. The artificial data set is based on an algebraic functional form, the Coulomb equation. The real-world data set is from a social science survey and deals with the influence of selected personal factors on subjective well-being. The difference in the structure of the data sets is in the number of input variables, as well as the complexity of the functional form and the correlations between the input variables. Moreover, the real data set includes a high but unknown level of noise.

The Online Supplement contains the data, the Keras/Python code for constructing and training the deep networks, and the Python code for using the hypothesis-testing framework.

### 4.1 First Example: Coulomb's Law Equation

This first data set depicts the behavior of a rather simple functional form. Coulomb's inverse square law – also known as Coulomb's law equation – describes the electrostatic force $F$ between two charged objects. The main features of the electrostatic force are the existence of two types of charges $q_1$ and $q_2$, the fact that like charges repel and unlike charges attract, and the decrease of force with separation $r$ between the two objects. The underlying mathematical formula was proposed by French physicist Charles Coulomb (1736-1806):

$$F = k\,\frac{q_1\,q_2}{r^2} \quad \text{with} \quad k = \frac{1}{4\,\pi\,\varepsilon\,8.854\cdot 10^{-12}}.$$

In this formula, $k$ is the electric force (Coulomb's) constant. It is dependent on the relative permittivity $\varepsilon$ of the medium, for example 1 for vacuum, 9.08 for the organochlorine compound Dichloromethane $CH_2Cl_2$, and 80.1 for water $H_2O$ (at a temperature of 20º C).

I create a random data set of about 125,000 tuples $\in [0; 1]$. I then scale the data so that $q_1, q_2 \in [0; 10]$; $r \in [0; 1.5]$; and $\varepsilon \in [1; 80]$. The calculated output variable falls into the range of $F \in [0; 1.931]$. Because the data follows an algebraic functional form, I can immediately write up four hypotheses:

**H1**: An increase of the charge $q_1$ is associated with an increase in the electrostatic force $F$ between the two charges.

**H2**: An increase of the charge $q_2$ is associated with an increase in the electrostatic force $F$ between the two charges.

**H3**: More separation $r$ between the two charges is associated with a decrease in the electrostatic force $F$ between the two charges.

**H4**: A larger permittivity $\varepsilon$ is associated with a decrease in the electrostatic force $F$ between the two charges.

4.1.1 Computational aspects of the deep network

To test these hypotheses, I build a deep artificial neural network with five layers: The first (input) layer has four nodes and the fifth (output) layer one node, matching the data set. I preprocess the variables so that they have a mean value of zero and a standard deviation of one. There are three hidden layers with fifty nodes each. For the input and hidden layers, the rectified linear unit (ReLU) serves as an activation function with a He normal initialization for the weights [54]. I initialize the weights leading to the output layer with the Glorot uniform technique [34]. Because this is ultimately a scalar regression problem, I match the network architecture with a mean squared error loss function [55] and use the efficient ADAM (adaptive moment estimation) optimization algorithm [56]. During training, I randomly remove a subset of nodes from the network with a dropout rate of 0.001 [57]. After dropout, I impose a weight constraint of five on the remaining nodes [58]. With these hyperparameters, fifty epochs are sufficient to train the network.

4.1.2 Results

On this trained network, I run the hypothesis-testing framework with a subset of 10,000 randomly selected tuples from the data set and $B = 50$ permutations. The results are summarized in Table 3.

The $R^2$ of the full model is 0.968. Charge $q_1$ exerts a large monotonic effect on the electrostatic force $F$, which is positive and statistically significant, Mann-Kendall $p < 0.001$; $f^2 = 8.210$; Theil-Senn $s = 0.007; p < 0.001$. As expected, the effect of charge $q_2$ is similar, Mann-Kendall $p < 0.001$; $f^2 = 8.115$;

Theil-Senn $s = 0.007; p < 0.001$. This supports H1 and H2. In support of H3, the separation $r$ exerts a large monotonic effect on $F$, which is negative and statistically significant, Mann-Kendall $p < 0.001$; $f^2 = 9.629$; Theil-Senn $s = -0.006; p < 0.001$. Backing H4, the permittivity $\varepsilon$ exerts a large monotonic effect on $F$, which is negative and statistically significant, Mann-Kendall $p < 0.001$; $f^2 = 23.166$; Theil-Senn $s = -0.004; p < 0.001$.

**Table 3: Variable effects (Coulomb's law)**

| Variable | $f_V^2$ | Monotonicity (Mann-Kendall) | P | Theil-Sen slope | p |
|---|---|---|---|---|---|
| Charge $q_1$ | 8.210 | Increasing | < 0.001 | 7.387·10⁻³ | < 0.001 |
| Charge $q_2$ | 8.115 | Increasing | < 0.001 | 7.772·10⁻³ | < 0.001 |
| Separation $r$ | 9.629 | Decreasing | < 0.001 | -6.140·10⁻³ | < 0.001 |
| Permittivity $\varepsilon$ | 23.166 | Decreasing | < 0.001 | -4.296·10⁻³ | < 0.001 |

This table reports the results of the hypothesis testing framework for example 1, Coulomb's law. The estimation is done on 10,000 randomly selected data records (about 8% of the entire data set) with $B = 50$ permutations; $R^2 = 0.965$.

**Figure 1: Accumulated local profiles (Coulomb's law)**

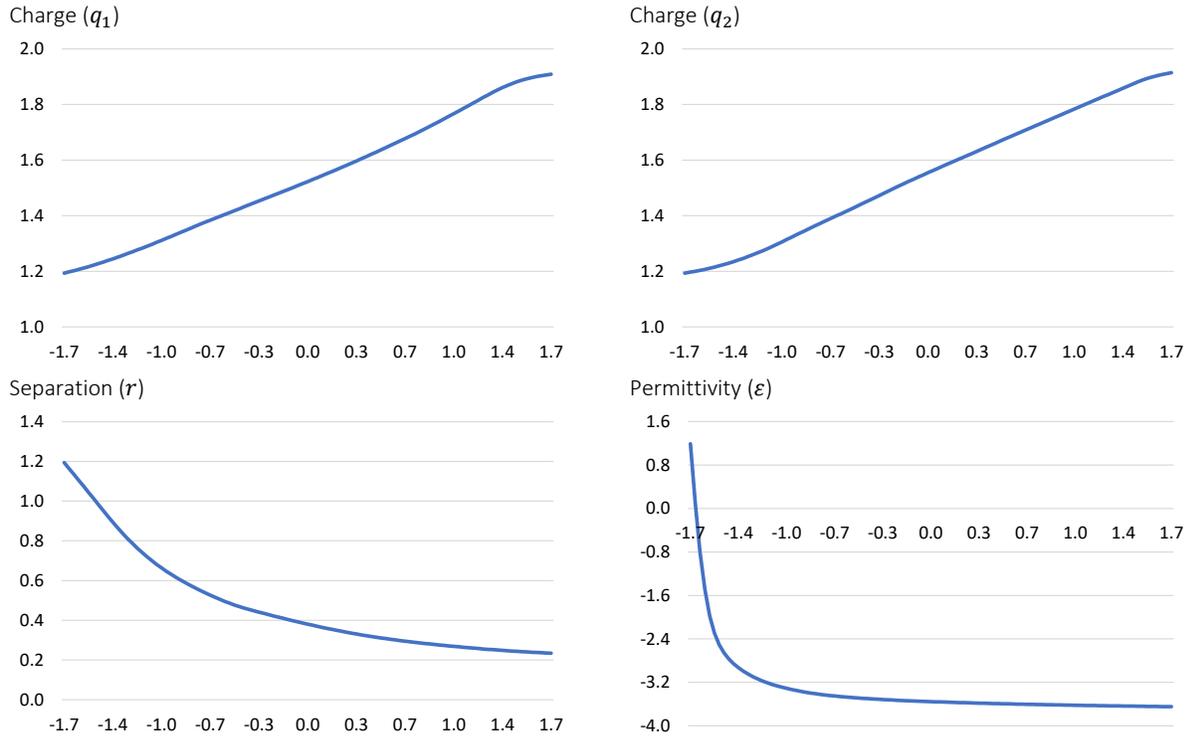

These plots show the accumulated local profiles for the variables used in example 1, Coulomb's law. The estimation is done on 10,000 randomly selected data records (about 8% of the entire data set) with $B = 50$ permutations.

A visualization of the variables' accumulated local profiles in Figure 1 provides additional confirmation. The profiles of $q_1$ and $q_2$ are practically straight-line, which is a sign of their linear relationships in Coulomb's law. The profiles of *r* and *ε* are monotonic hyperbolae, which reflects them being denominators in Coulomb's law.

## 4.2 Second Example: Subjective Well-Being

The second data set consists of 25,700 responses to a social science survey administered in Germany, the European Social Survey [ESS; 59]. I am concerned with how personal value priorities, opinions, and other individual factors affect personal well-being [60,61]. Because well-being is an emotional mental state created out of the combination of basic psychological operations and ingredients, which can be mapped to intrinsic networks in the human brain [62], a representation with a deep learning artificial neural network is especially appropriate.

Various forms of well-being are empirically separate [63], and thus a delineation of the concept as I use it here is important. Following past research into well-being [e.g., 64,65] as well as pertinent applied literature [e.g., 66,67], I view subjective well-being as including both cognitive and affective components. The cognitive component is someone's overall life satisfaction, and the affective component is the difference in how frequent positive and negative emotions are experienced. Taken together, the cognitive and affective component commonly equate with what is called everyday happiness and satisfaction [65,68]. The ESS measures cognitive and affective well-being with two items on a 11-point Likert-scale. Following guidelines for happiness research [67,69], I compute the output variable subjective well-being as a simple arithmetic average of these two items.

As input variables, I select 159 variables from the ESS describing the individual (see the detailed variable list in the Online Supplement [70]). For my exemplary hypotheses, I select four classic variables that previous happiness research has shown to be related to subjective well-being [71]. Income and education differences between individuals are usually correlated with reports of well-being, but previous research has shown that such positional differences explain no more than 10% of the variance in subjective well-being [72]. The ESS variable *hincfel* asks for the respondent's feeling about the household income nowadays (4: very difficult; 3: difficult; 2: coping; and 1: living comfortably on present income), and *eduyrs* contains the number of years of education completed. Well-being includes positive elements that transcend economic prosperity [73], with health perhaps the most important of all dimensions shaping the quality of life [74]. The common variance explained by variables related to life ability is usually around 30% [72]. In this vein, the variable *health* reflects a respondent's subjective health in general (5: very bad to 1: very good). Trust is an essential element in any social setting and drives economic effects, health, and well-being [e.g., 75,76]. The variable *ppltrst* asks respondents if most people can be trusted, or if one cannot be too careful in dealing with people (0: you can't be too careful; 10: most people can be trusted).

Taken together, this leads to the following hypotheses:
**H1**: Higher income is associated with higher levels of subjective well-being.
**H2**: Higher levels of education are associated with higher levels of subjective well-being.
**H3**: Better general health is associated with higher levels of subjective well-being.
**H4**: More trust in other people is associated with higher levels of subjective well-being.

### 4.2.1 Computational aspects of the deep network

To test the hypotheses, I build a deep network with 159 nodes in the input layer describing an individual survey respondent, one node in the output layer representing this person's well-being, and four hidden layers with 500 nodes each. I transform categorical variables into binary placeholder variables (see Online Supplement) and rescale all other variables to an interval of [0; 1]. I fill occasional missing values with the k-Nearest Neighbors approach (*k* = 5; points weighted by the inverse of their distance). I preprocess all input and output data to give a mean value of zero and a standard deviation of one. For the input and hidden layers, the ADAM optimization algorithm together with a ReLU activation function with He normal initialization is used. For the output layer, a linear function with a Glorot uniform weight initialization is deployed. To improve the model's generalization capability for small data sets, I add a L$_2$ regularizer with a very small weight decay parameter of λ = 0.001 to each layer and the bias [77]. During training, I randomly remove a subset of nodes from the network with a dropout rate of 0.1; I impose a weight constraint of three to the remaining nodes after the dropout. Because the variables are all derived from answers to Likert-type survey questions, I add a zero-centered Gaussian Noise (standard deviation of 0.01) as random data augmentation to the input and output layers. This aids generalization and fault tolerance through embodying a smoothness assumption [33,78]. Ten epochs are sufficient to train the network.

### 4.2.2 Results

On this trained network, I run the hypothesis-testing framework with a subset of 10,000 randomly selected tuples from the data set and 50 permutations. The results are summarized in Table 4.

The $R^2$ of the full model is 0.527. Variable *hincfel* exerts a small monotonic effect on *well-being*, which is negative and statistically significant, Mann-Kendall $p < 0.001$; $f^2 = 0.116$; Theil-Senn $s = 0.006$; $p < 0.001$. Because *hincfel* is reverse scored, this supports H1, that is, higher income is generally associated with higher levels of subjective well-being. In support of H2, longer time spent on education (*eduyrs*) has a negative, monotonic and statistically significant effect on subjective well-being, Mann-Kendall $p < 0.001$; $f^2 = 0.004$; Theil-Senn $s = 0.001$; $p < 0.001$. Because of the extremely small magnitude of the effect, this relationship is trivial and practically not relevant. "Although not unrelated, the size and statistical significance of effects are logically independent features of data from samples" [3]. The variable *health* shows a small effect on well-being, which is negative and statistically significant, Mann-Kendall $p < 0.001$; $f^2 = 0.079$; Theil-Senn $s = 0.005$; $p < 0.001$. Because *health* is reverse scored, this supports H3. When people trust others (*ppltrst*), this has a monotonic effect on well-being, which is, in support of H4, positive and statistically significant, Mann-Kendall $p < 0.001$; $f^2 = 0.006$; Theil-Senn $s = 0.136 \cdot 10^{-3}$; $p < 0.001$. Similar to the variable *eduyrs*, this effect is trivial and does not have practical significance.

A visualization of the variables' accumulated local profiles in Figure 2 provides additional confirmation. The profile of *health* clearly shows a linear relationship. The profile of *hincfel* is slightly bent downwards (note again that *hincfel* is reverse scored), which shows that people with high-income report disproportionally higher well-being.

**Table 4: Variable effects (subjective well-being)**

| Variable | $f_V^2$ | Monotonicity (Mann-Kendall) | P | Theil-Sen slope | p |
|---|---|---|---|---|---|
| Income (*hincfel*) | 0.116 | Decreasing | < 0.001 | -6.562·10⁻³ | < 0.001 |
| Education (*eduyrs*) | 0.004 | Decreasing | < 0.001 | -1.510·10⁻³ | < 0.001 |
| Health (*health*) | 0.079 | Decreasing | < 0.001 | -5.827·10⁻³ | < 0.001 |
| Trust (*ppltrust*) | 0.006 | Increasing | < 0.001 | 0.136·10⁻³ | < 0.001 |

This table reports the results of the hypothesis testing framework for example 2, subjective well-being. The estimation is done on 10,000 randomly selected data records (about 39% of the entire data set) with $B = 50$ permutations; $R^2 = 0.527$.

**Figure 2: Accumulated local profiles (subjective well-being)**

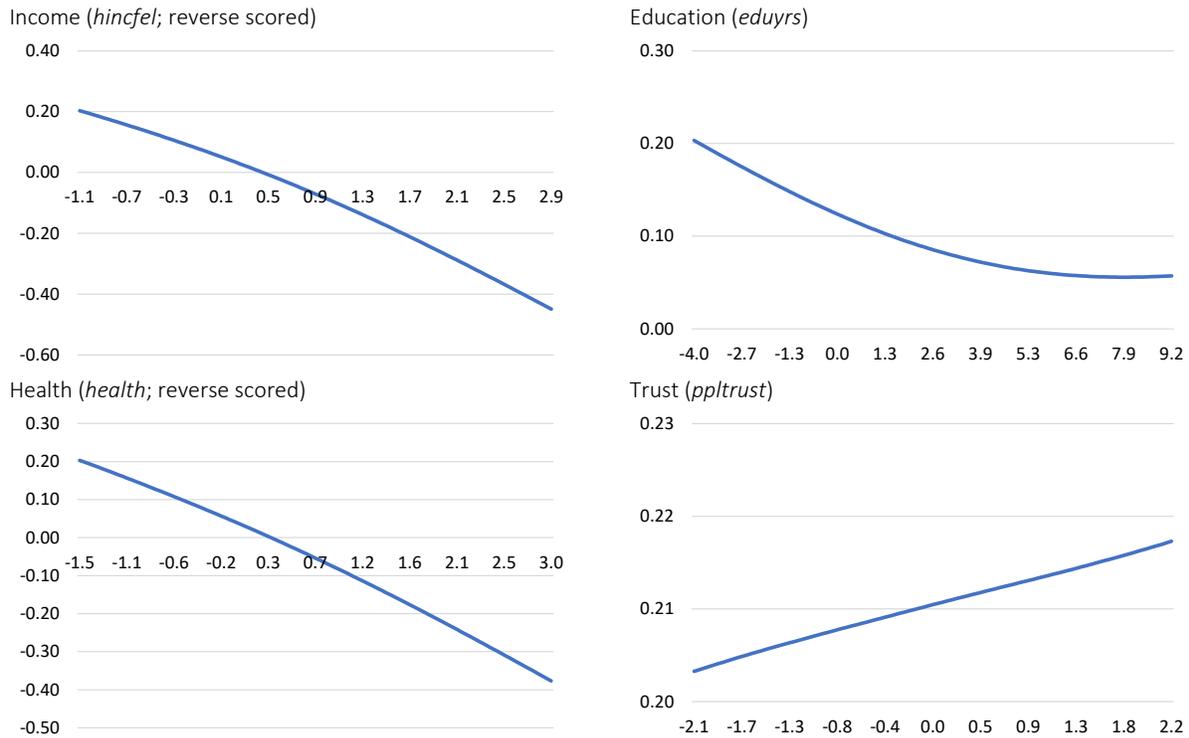

These plots show the accumulated local profiles for the variables used in example 2, subjective well-being. The estimation is done on 10,000 randomly selected data records (about 8% of the entire data set) with $B = 50$ permutations.

## 5. Discussion

Machine learning models are not commonly used by researchers for hypothesis testing. While they would be much better suited to address many interesting research questions about complex statistical relationships than multivariate OLS regression models, they hide their statistical reasoning in a black box. XAI has developed techniques to assess the connection between input and output variables with respect to strength, direction, and form. While these novel XAI methods are mainly geared towards distinguishing important variables from not so important ones using visualization, they do not relate well to classical measures of inferential statistics. In fact, "in machine learning, the concept of interpretability is both important and slippery" [79]. Till date, this has unfortunately weakened the acceptance of machine learning models for hypothesis testing in scientific experiments and research models.

In this paper, I have outlined an approach to estimate Cohen's $f^2$ for the input variables of a deep learning network and to show their direction of influence and statistical significance. This now allows researchers to move beyond multivariate OLS regression and analyze regression problems with more complex deep learning networks – using and reporting information they are familiar with from inferential statistics. In the context of machine learning, it has been shown "that explanations fare better when they require less cognitive effort" [80] on behalf of the researcher. To aid my approach, I present Python software, which builds upon Keras, a deep learning open-source software library, and extends dalex, a model-agnostic explainer interface.

While I have tested my hypothesis-testing framework on several data sets and documented two examples in this paper, there is more research to be done:

First, estimating the effect size with a variable permutation algorithm is not necessarily equivalent to the traditional loco approach. As I have explained in the article, the latter is not an option for machine learning. I have suggested steps to adjust the effect size calculated by the former. On the data sets that I have used during my tests, the effect size thresholds proposed by Cohen seem to be reasonable. But, clearly, this calls for further corroboration.

Second, when a machine learning model performs very poorly, the model's $R^2$ can have a negative value [81]. This just indicates that the model performed poorly, but it is impossible to know how bad exactly it performed, because the lower bound for $R^2$ is $-\infty$. While hypothesis testing of a variable's effect in such a model is obviously pointless, my approach for calculating effect sizes does not explicitly take care of that (rare) situation.

Third, to summarize the influence of an independent variable on the dependent variable, I have used a Theil-Sen estimator on the variables' accumulated local effect profile. I have argued that the accumulated local effect profile is better suited than the local dependence profile or partial dependence plot because it does not contain correlated effects of other independent variables. I have converted the slope's confidence interval into a *p*-value for adjudicating on the slope's statistical significance. There is further investigation and experimentation to be done to confirm that approach.

Fourth, in the context of neural networks, an independent variable is rarely important on its own [82]. Notwithstanding, I have not yet considered interactions within a group of independent variables. In multivariate OLS regression models, this frequently leads to an issue of multicollinearity. The redundant architecture of deep artificial neural networks, however, allows to learn a potential similarity of input variables and predict even in situations of multicollinearity [83–85]. Thus, a potential further development of the hypothesis-testing framework should address interactions between independent variables.

## Online Supplement: Code and Data Availability

The Python code for the hypothesis-testing framework, the worked examples, and the data sets are available in the online repository OSF.IO (DOI: 10.17605/OSF.IO/QDJCY).